\documentstyle[psfig]{mn}
\def\H0{{\it H}$_0$}

\def\q0{{\it q}$_0$}

\def\ergps{erg~s$^{-1}$}
\def\ergpMpc{erg~Mpc$^{-3}$}

\def\ni{\noindent}

\def\spose#1{\hbox to 0pt{#1\hss}}
\def\approxlt{\mathrel{\spose{\lower 3pt\hbox{$\sim$}}
        \raise 2.0pt\hbox{$<$}}}
\def\approxgt{\mathrel{\spose{\lower 3pt\hbox{$\sim$}}
        \raise 2.0pt\hbox{$>$}}}
    
\title[High redshift extended X-ray emission: radio galaxies vs.
clusters] {Extended X-ray emission at high redshifts: radio galaxies
versus clusters}
\author[A. Celotti \& A.C. Fabian] {\parbox[]{6.5in} 
{A. Celotti$^1$ and A.C. Fabian$^2$}\\ \\
$^1$SISSA, via Beirut, 2-4, 34014 Trieste, Italy\\ $^2$Institute of
Astronomy, Madingley Road, Cambridge CB3 0HA\\}

\date{}

\begin{document}

\maketitle

\begin{abstract}
Most old distant radio galaxies should be extended X--ray sources due
to inverse Compton scattering of Cosmic Microwave Background (CMB)
photons. Such sources can be an important component in X-ray surveys
for high redshift clusters, due to the increase with redshift of both
the CMB energy density and the radio source number density. We
estimate a lower limit to the space density of such sources and show
that inverse Compton scattered emission may dominate above redshifts
of one and X-ray luminosities of $10^{44}$\ergps, with a space density
of radio galaxies $> 10^{-8}$ Mpc$^{-3}$. The X-ray sources may last
longer than the radio emission and so need not be associated with what
is seen to be a currently active radio galaxy.
\end{abstract}

\begin{keywords}galaxies: active - galaxies: clusters - X-ray: galaxies - clusters
\end{keywords}

\section{Introduction}
{\em Chandra} has revealed extended X-ray emission from a wide range
of radio sources out to high redshifts. Jets and the lobes and cocoons
of radio quasars and galaxies have been imaged with unprecedented
resolution (e.g. Chartas et al 2000, Schwarz et al 2001, Harris \&
Krawczynski 2002, Kataoka et al 2003, Comastri et al 2003, Wilson,
Young \& Shopbell 2001, Kraft et al 2002). At low redshifts, extended
emission directly associated with the radio lobes is seen through its
inverse Compton emission in some objects (e.g. Fornax A; Feigelson et
al 1995, Kaneda et al 1995; 3C219, Comastri et al 2003). In the
powerful radio source Cygnus A, diffuse X-ray emission is also
detected from the radio cocoon, i.e.  the reservoirs of shocked
material associated with the radio expansion (Wilson, Young \& Smith
2003). When however a modest radio source lies in a rich cluster, the
surface brightness of the inverse Compton emission in soft X-rays can
be so low that it cannot be separated from thermal emission and the
lobes appear as holes in the X-ray emission due to displacement of the
hot gas by the lobes (e.g. the Perseus cluster, Fabian et al 2000,
Sanders et al 2004; Hydra A, McNamara et al 2000; A2052, Blanton et al
2001).

Extended X-ray emission is also associated with an increasing number
of radio sources at cosmological distances, for example 3C~294 (Fabian
et al 2003), 3C~9 (Fabian, Celotti \& Johnstone 2003), PKS\,1138-262
(Carilli et al 2002), 4C~41.17 (Scharf et al 2003), GB 1508+5714 (Yuan
et al 2003, Siemiginowska et al 2003), although often the low photon
count rate makes it hard to disentangle the different X-ray
components. In general however, much of the emission can be
interpreted as due to inverse Compton scattering of non-thermal radio
emitting electrons on CMB photons (Felten \& Rees 1967, Cooke,
Lawrence \& Perola 1978, Harris \& Grindlay 1979). Where directly
associated with powerful jets relativistic bulk motion may be involved
(Celotti, Ghisellini \& Chiaberge 2001; Tavecchio et al 2000). The
steep increase in the energy density of the CMB with redshift $z$ (as
$(1+z)^4$) partially compensates for the large distance to such
sources (Felten \& Rees 1967, Schwartz 2002), thereby making them
detectable.

Independently of the spatial distribution, the presence of extended
radio synchrotron emission is a direct indication of the presence of a
non-thermal population of relativistic particles. These particles, at
least, must produce high energy emission via inverse Compton
scattering of CMB photons (direct measurements and upper limits for
such emission are used to estimate the intracluster magnetic field).
Here, we consider the effective number density of extended X-ray
emitting sources due to this process as a function of X-ray luminosity
and show that they can be a serious contaminant to X--ray surveys
searching for clusters and protoclusters at high redshift.

The outline of the paper is as follows: in Section 2 we estimate
the ratio of (synchrotron) radio to (inverse Compton) X-ray emission,
while in Section 3 we estimate the corresponding X-ray luminosity
functions of radio sources as a function of $z$ and compare them with
those of X-ray clusters. A discussion in Section 4 concludes this
Letter.  A cosmology with $\Omega_{\Lambda}=0.7, \Omega_{\rm M}=0.3$,
$H_0=50$ km s$^{-1}$ Mpc$^{-1}$ has been assumed.

\section{X-ray emission from scattering on the CMB}

We now attempt to estimate the X-ray luminosity associated with each
extended radio source and the number density of X-ray emitting objects
using radio emission as a direct tracer of the relativistic electron
population. Although based on well known classical arguments (Felten
\& Morrison 1966), we explicitly re-derive the limits on the inverse
Compton emission as these are the basis for the robustness of the
inferred X--ray emission from radio sources.  A simplified approach
restricts us to lower limits, but bypasses the need for determining
particle life-times (due to acceleration/injection history and
radiative and adiabatic cooling, see Sarazin 1999) and consideration
of the (uncertain) conversion of radio luminosity into source
power. It should be noted that for typical gas densities associated
with radio emitting regions, inverse Compton largely dominates over
non-thermal bremsstrahlung emission (Sarazin 1999, Petrosian 2001).

In order to include only radio emission arising as synchrotron
radiation from a non-thermal (power-law) distribution of electrons on
extended scales we use low frequency radio luminosities, $L_{\rm R}$,
at 151 MHz.  The major uncertainty in the estimate of the X-ray
luminosity is the magnetic field in the radio-emitting
region. Depending on its intensity the same electrons can give both
the observed radio emission and the X-ray emission, or some
extrapolation of the electron spectrum is required.

Let us consider the monochromatic X-ray luminosity $L_{\rm x}$ at a
reference frequency of 1 keV (in the observer frame), and define
$\gamma_{\rm x}= 3 \nu_{\rm x} /4 \nu_{\rm CMB}\simeq 10^3$ as the
Lorentz factor of the electrons which could emit at (the observed)
1~keV photon frequency $\nu_{\rm x}$ via inverse Compton on the CMB
(peaked at the frequency $\nu_{\rm CMB}$) and $B_{\rm x}\sim$ few
$\times 10^{-5}$ as the magnetic field intensity for which these same
electrons radiate at $\nu_{\rm R}$ (151 MHz) via synchrotron
(i.e. $B_{\rm x}\equiv 5.4\times 10^{-5} \nu_{\rm R} (1+z) \gamma_{\rm
x}^{-2}$ G for $z=0$).  In other words, for any magnetic field $B <
B_{\rm x}$, the Lorentz factor of the radio (151 MHz) emitting
particles $\gamma_{\rm R} > \gamma_{\rm x}$, and vice-versa.

The relative luminosity in the synchrotron and inverse Compton
components is given by the ratio of the magnetic $U_{\rm B}$ vs
radiation $U_{\rm CMB}$ energy densities. More specifically, the
relative luminosities at the two fixed observed frequencies in the
radio and X-ray bands scale as

\begin{equation}
\frac{\nu_{\rm x} L_{\rm x}}{\nu_{\rm R} L_{\rm R}} = \frac{U_{\rm
CMB}}{U_{\rm B}} \left(\frac{\nu_{\rm X}\, \nu_{\rm B}} {\nu_{\rm R}\,
\nu_{\rm CMB}}\right)^{1-\alpha} (1+z)^{(3+\alpha)-k(1+\alpha)},
\end{equation}
where $U_{\rm B}$ and $U_{\rm CMB}$ are the energy densities at
redshift $z=0$ and $k$ accounts for a possible dependence of the
magnetic energy density on $z$, parametrized as $B(z)=B(0) (1+z)^k$.
The non-thermal particle distribution has been assumed to be a
power-law whose slope $p$ is related to the luminosity spectral index
$\alpha = (p-1)/2$ ($L(\nu)\propto \nu^{-\alpha}$).

\begin{figure}
\psfig{figure=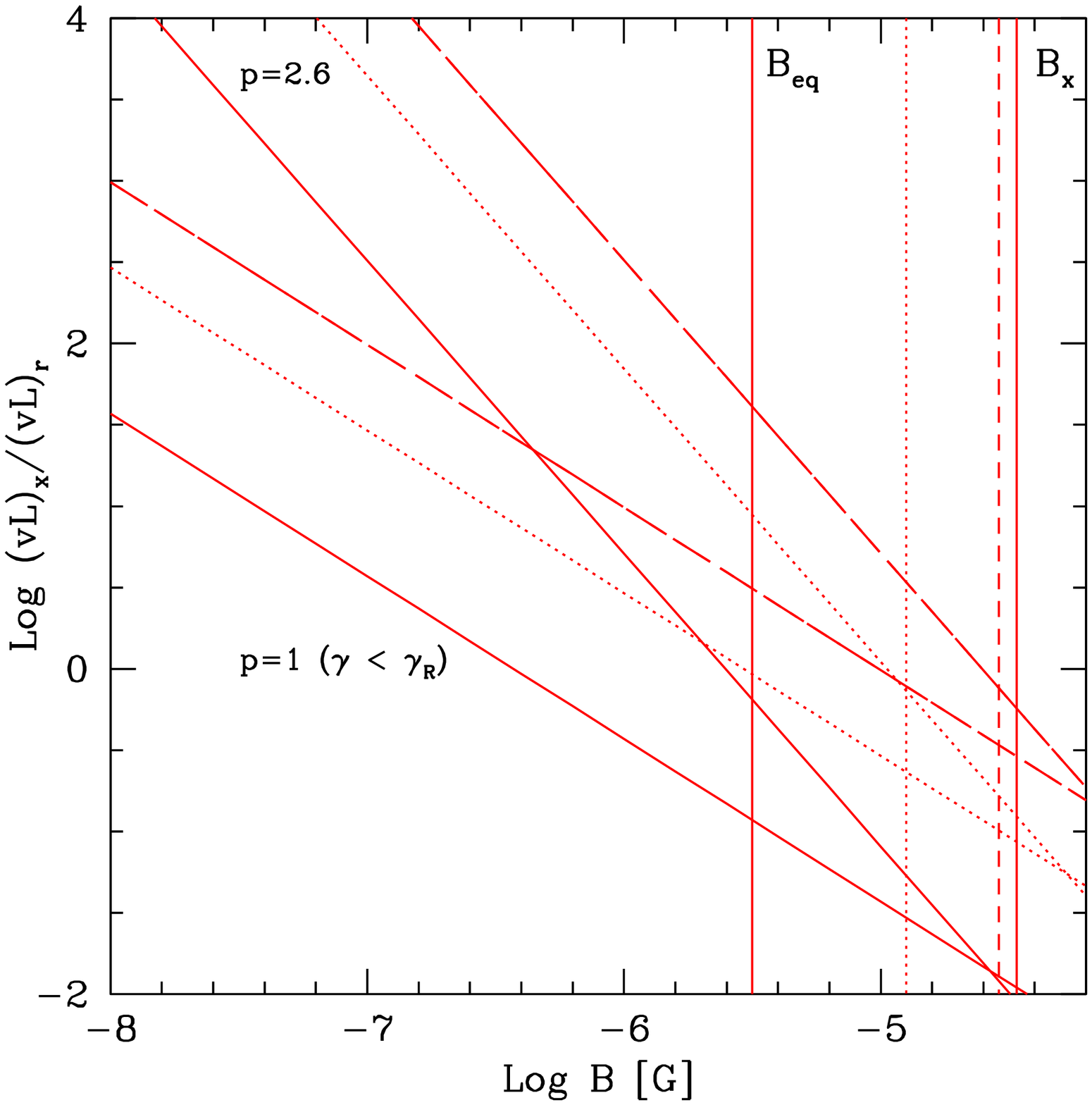,width=0.45\textwidth} 
\psfig{figure=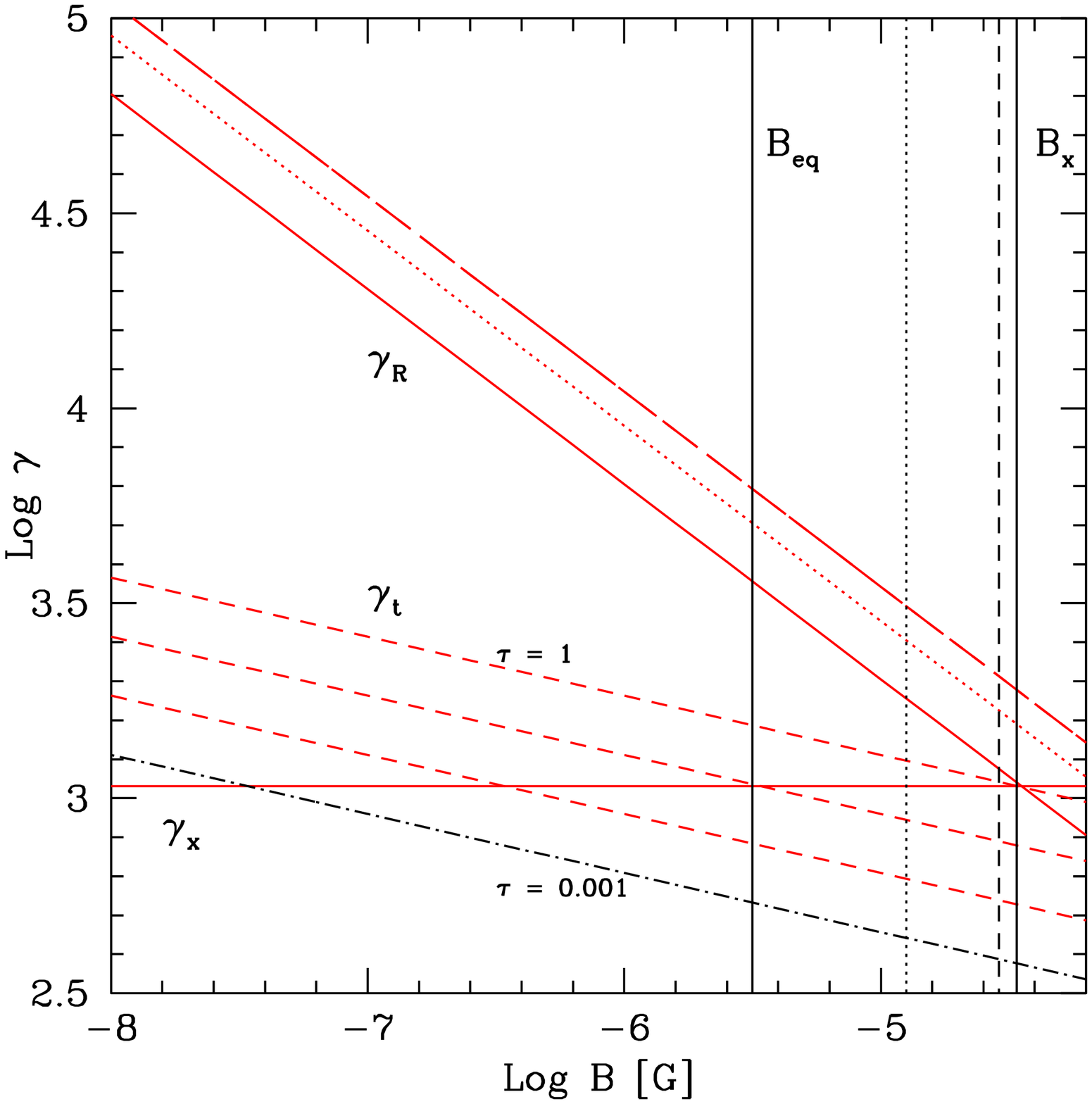,width=0.45\textwidth}
\caption{ In both panels $B_{\rm eq}$ is the value of the field in
  equipartition with $U_{\rm CMB}$ (at redshifts z=0,1,2, - solid,
  dotted and dashed lines, respectively).  The solid vertical line
  labeled $B_{\rm x}$ represents the value of $B$ for which electrons
  emitting at 1 keV via inverse Compton radiate at 151 MHz via
  synchrotron. Top panel: ratio of the expected X-ray (inverse
  Compton) and the radio (synchrotron) luminosities at the observed
  frequencies of 1 keV and 151 MHz for a single power-law particle
  distributions with $p$=2.6, and one flattening to $p=1$ at
  $\gamma<\gamma_{\rm R}$. The luminosity ratio is represented by the
  oblique lines (solid, dotted and dashed for $z=0,1,2$,
  respectively).  Bottom panel: the Lorentz factors of the emitting
  electrons (for the assumed $B$ field).  The horizontal lines
  indicate $\gamma_{\rm x}$, i.e. the Lorentz factor of the electrons
  which could emit at (the observed) 1 keV energy via inverse Compton
  on the CMB, while the oblique lines (labeled $\gamma_{\rm R}$) show
  the Lorentz factors of electrons emitting at 151 MHz (for z=0,1,2 -
  solid, dotted and dashed lines, respectively). The lines labeled
  $\gamma_{\rm t}$ indicate instead the synchrotron self-absorption
  frequency for different (Thomson) optical depths ($\tau= 10^{-3},
  10^{-2}, 10^{-1}, 1$). It is assumed a homogeneous region and
  constant $B$ as a function of redshift.}
\end{figure}

Figs. 1a and 2a show the ratio of the expected X-ray (inverse Compton)
and radio (synchrotron) luminosities at the (observed) frequencies of
1 keV and 151 MHz, at different redshifts and for two representative
power-law particle distributions, with $p$=2.6 and 2, respectively.
Reference values for the magnetic field ($B_{\rm eq}$, i.e. the field
in equipartition with $U_{\rm CMB}$, and $B_{\rm x}$) are also
reported as vertical lines.  The X-ray/radio luminosity ratio ranges
(case $p=2.6$) for $0<z<2$, between $300-3\times 10^4$ (8$\times
10^{-2}- 8$) for a $0.1\mu$G (10$\mu$G) field.  [For $p=2$, analogous
ranges imply ratios spanning $50-3\times 10^3$ for $B\sim 0.1\mu$G
($8\times 10^{-2}-3$ for $B\sim 10\mu$G)]. Alternatively, for fields
in the interval 1-10$\mu$G and $z>1$, $L_{\rm x} \nu_{\rm x}/ L_{\rm
R} \nu_{\rm R}>0.5$.  These estimates assume a homogeneous region and
$B=$ constant as a function of redshift (i.e. $k=0$).

\subsubsection{Caveats}

There are two main caveats to be considered before adopting the above
estimates.

Firstly, the actual value of the magnetic field. Estimates for nearby
clusters based on equipartition (minimum energy) arguments from the
non-thermal emitting component lead typically to $B\sim 0.1-1 \mu$G
for unit volume filling factor and equal electron and proton energies.
Similar estimates are derived for the few clusters where there is a
detection of hard X-ray emission \footnote{see however the recent
results by Rossetti \& Molendi (2003) on the Coma cluster.}, assumed
to be inverse Compton scattering on CMB photons. Fields inferred from
Faraday rotation measures are instead about one order of magnitude
higher (including cooling flow clusters), in the range $1-10\mu$G
(Carilli \& Taylor 2002, Feretti 2003 for reviews). These values can
be reconciled with the former ones by taking into account the
dependence on radial distance, field substructures, different electron
spectra, etc. (see e.g. Feretti 2003).  In any case, typical fields
appear to be in the range considered above, up to $\sim 10\mu$G,
i.e. $\approxlt B_{\rm x}$, and thus not dynamically important in
clusters.  Although the origin of such fields is not known, it appears
likely that seed fields (primordial or produced from galactic winds,
AGN, shocks associated to large scale structure formation) could be
amplified following cluster mergers up to $\sim \mu$G values
(e.g. Roettiger, Stone \& Burns 1999).  Therefore, it seems plausible
that, despite a higher (thermal) gas pressure at higher cluster
luminosities/redshift, the actual field might be a decreasing function
of $z$, leading to even higher $L_{\rm x} \nu_{\rm x}/ L_{\rm R}
\nu_{\rm R}$ values. We conclude that the above estimates for $B\sim
1-10\mu < B_{\rm x}$ G fields are a reasonably robust lower limit to
such ratio.

The second critical assumption in the above estimates is the shape of
the particle distribution, considered as a single power-law.  As this
might well not be the case, let us thus consider the uncertainties
related to this assumption, which depend on which energy range of the
particle distribution is actually responsible for the X-ray emission.
Figs. 1b and 2b show the Lorentz factors of the emitting electrons as
a function of $B$.  The horizontal line indicates $\gamma_{\rm x}$
while the oblique lines (labeled $\gamma_{\rm R}$) show the Lorentz
factors of electrons emitting at 151 MHz at different redshifts.

For any $B< B_{\rm x}$ (as discussed above), corresponding to
$\gamma_{\rm R}> \gamma_{\rm x}$, it is therefore necessary to assess
whether the extrapolation on the particle number density at energy
$m_{\rm e} c^2 \gamma_{\rm x}$ from the corresponding intensity at 151
MHz provides a robust lower limit, as the particle distribution might
flatten or cut off at a Lorentz factor $\gamma^*$, with $\gamma_{\rm
x}< \gamma^* <\gamma_{\rm R}$.  Indeed such a flattening/cutoff could
be expected if:

\ni a) relativistic particles have been injected with energies
$>\gamma_{\rm x}$ and had not (yet) cooled down to $\gamma_{\rm x}$;
the observed spectral slope would then correspond to the slope of a
cooled particle distribution, i.e. above a cooling break $\gamma_{\rm
b}$ with $\gamma_{\rm R} > \gamma_{\rm b}>\gamma_{\rm x}$; the low
energy particles lose energy non radiatively, namely via Coulomb
losses -- indeed for typical (cluster) densities this process
dominates at $\gamma< 100$ (e.g. Petrosian 2001);

\ni b) self-absorption is effective in re-heating the low energy
particles, causing the particle distribution to (quasi-) thermalize
around a self-absorption Lorentz factor $\gamma_{\rm t}>\gamma_{\rm
x}$ (Ghisellini, Guilbert \& Svensson 1988).  The `flattest' oblique
lines in Figs. 1b, 2b represent $\gamma_{\rm t}$ for increasing values
of the (Thomson) optical depth (for $\tau = 10^{-3}, 10^{-2}, 10^{-1},
1$), showing that indeed self-absorption might cause the particle
spectrum to flatten at energies higher than $\gamma_{\rm x}$, although
only for significantly large optical depths in relativistic particles.

The radiative cooling timescales of the X--ray emitting electrons
\begin{equation}
t_{\rm rad} \simeq 2.4\times 10^9 \gamma^{-1}_{\rm x} (1+z)^{-4} 
\left(1+\frac{U_{\rm B}}{U_{\rm CMB}} (1+z)^{k-4}\right) \quad {\rm yr}
\end{equation}
would be typically much larger than the adiabatic one 
\begin{equation}
t_{\rm ad}\simeq 3.2\times 10^6
R_2 (R_2/\lambda_{\rm scatt,1})\qquad\qquad {\rm yr}
\end{equation}
where $R_2 = R /100$ kpc is a typical size of extended radio emission,
and $\lambda_{\rm scatt}=10$ kpc is considered as the typical
`scattering' length for field coherence cells of $\sim 10$ kpc
(Carilli \& Taylor 2002).  Furthermore expansion -- and thus cooling -
would likely start from the injection/acceleration site, i.e. on
presumably smaller scales.  Thus it appears unlikely that particles
with $\gamma>\gamma_{\rm x}$ have not cooled or that particle
thermalisation could be effective (also inverse Compton cooling should
prevent the quasi-thermalization of the non-thermal particles below
$\sim \gamma_{\rm t}$). Note also that Coulomb losses start to
dominate over radiative ones for $\gamma < 100$, not affecting the
shape of the particle distribution above $\gamma_{\rm x}$.

In order to estimate the effect on the X-ray-to-radio luminosity ratio
of a break in the particle distribution we considered an `extreme'
case where the particle spectrum flattens to $p=1$ (i.e. as expected
from adiabatic cooling and flatter than for radiative cooling) just
below $\sim \gamma_{\rm R}$. The corresponding luminosity ratio is
shown in Fig. 1a: for $B < 10 \mu$G the ratio is $\approxgt 0.3$ at
$z\approxgt 1$.

It is worth noticing that these considerations also account for an
inhomogeneous radio emitting volume (equivalent to the added
contributions from different regions), as the above results would
refer to the X-ray luminosity expected from the region dominating the
radiative output at 151 MHz.  In this case however the observed
spectrum would not be indicative of the shape of the emitting particle
distribution: if the latter is steeper than $p=2.6$ the above
estimates (for $p=2.6$) provide a lower limit to $\nu_{\rm x} L_{\rm
x}/\nu_{\rm R} L_{\rm R}$. The reported estimates for $p=1$ provide
instead a lower limit for any flat distribution with $p\approxgt 1$.

\begin{figure}
  \psfig{figure=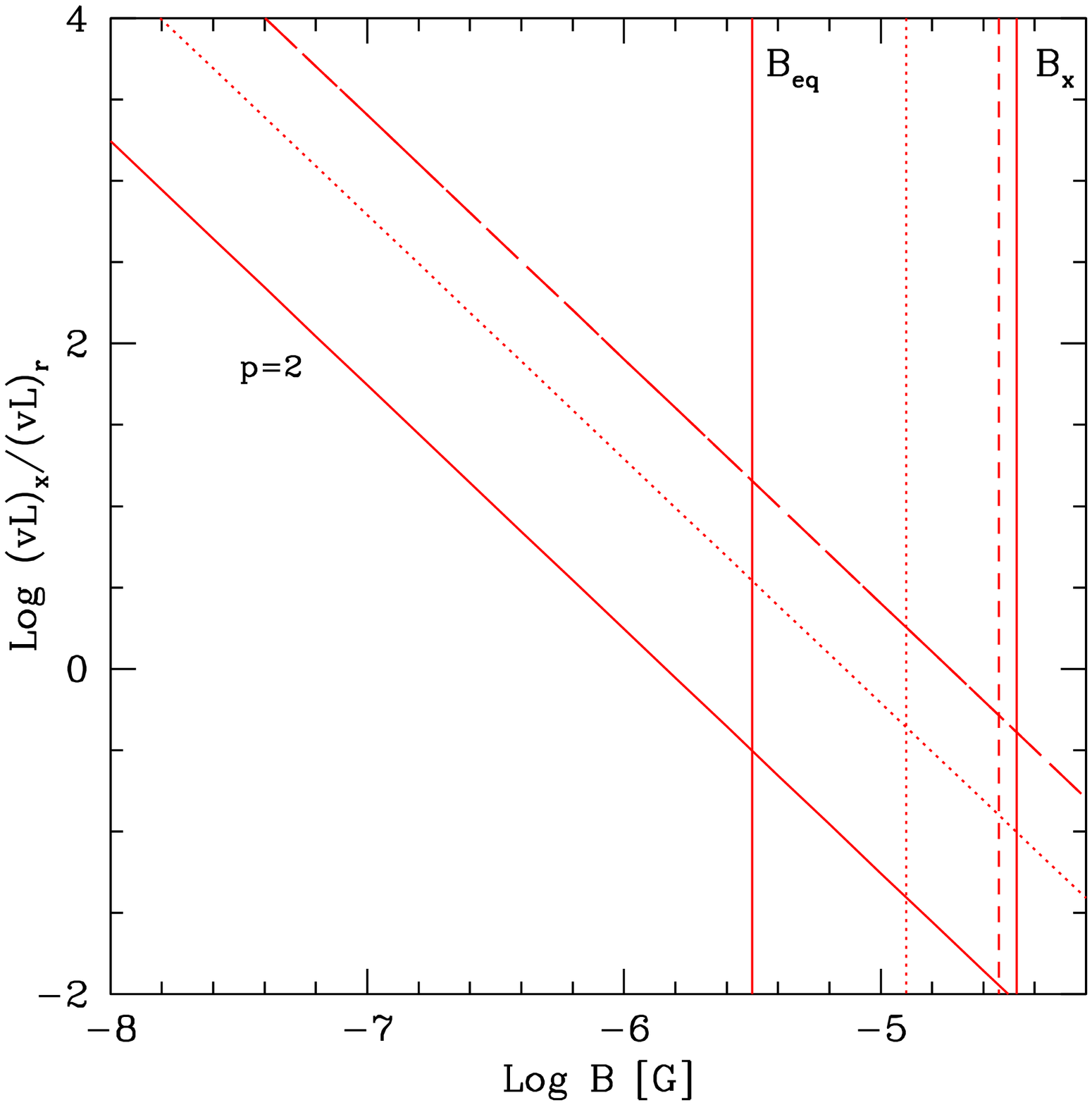,width=0.45\textwidth}
  \psfig{figure=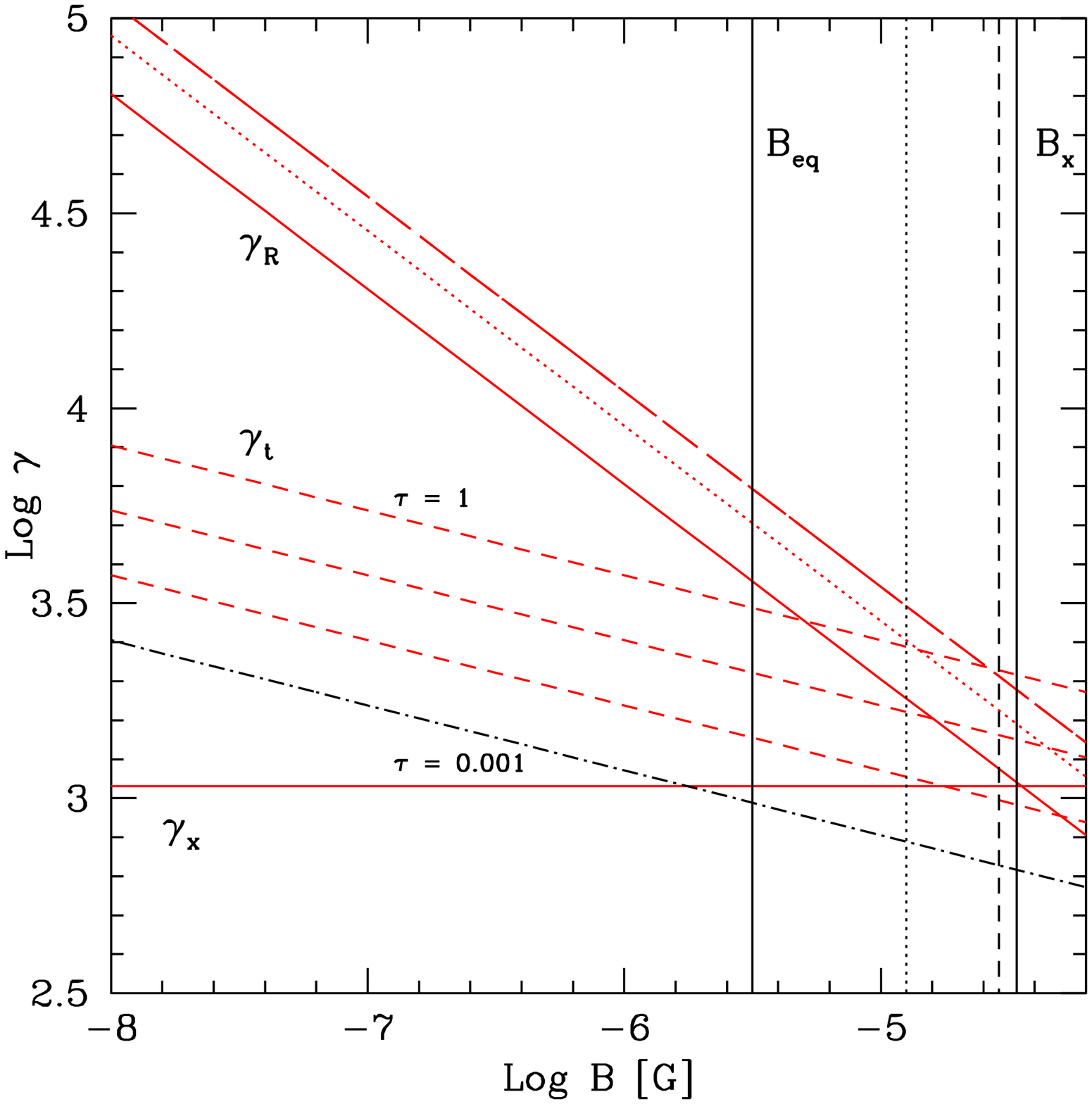,width=0.45\textwidth}
\caption{Same as Fig. 1, for a particle distribution with $p=2$.}
\end{figure}

We conclude that radio sources constitute a potentially significant
population of extended X-ray emitting objects.

\section{\bf High redshift extended X-ray emitters} 

In order to estimate the number density and thus the possible
contribution of radio sources to X-ray surveys (see also Schwartz
2002), we consider the luminosity function of steep radio sources, as
determined from surveys at 151 MHz. This has a twofold advantage: it
avoids or at least reduces contamination from compact components (with
respect to the extended lobe/cocoon emission) and it takes advantage
of the most recent and deeper radio surveys.

In particular we adopt the luminosity function and evolution
determined by Willott et al (2001) from the 3CRR, 6CE and 7CRS samples
\footnote{Their results have been converted here for the different
assumed cosmology.}. They parametrized (see also Dunlop \& Peacock
1990) the radio source luminosity function and its evolution as due to
two (distinct) populations: FR~I plus low excitation line (LEG) FR~II
sources and high excitation line (HEG) FR~II, representing low and
high radio power sources, respectively (Fanaroff \& Riley 1974, see
Laing et al 1994, Jackson \& Wall 1999). We associate an extended
X-ray luminosity with each 151~MHz radio sources as determined above:
Fig. 3 shows the corresponding number density of the two populations
of X-ray sources at different redshifts assuming a conservative (and
constant) radio-to-X-ray (151 MHz and 1 keV) conversion factor $L_{\rm
x} \nu_{\rm x}/L_{\rm R} \nu_{\rm R} = 1$.

For comparison, Fig. 3 also reports the luminosity function and
evolution of X-ray clusters in the (0.5-2) keV band following the
parametrization by Rosati et al (2000) -- although the evolution at
luminosities $<$ few$\times 10^{44}$ erg s$^{-1}$ is currently not
robustly determined by the data.  It is apparent that at redshift
$z\approxgt 1$ the radio source population starts becoming comparable
to or even exceeding the expected cluster number density at the high
X-ray luminosity end, $L_{\rm x} > 10^{44}$ erg s$^{-1}$.  Because of
the weaker cosmological evolution and intrinsic lower luminosity of
the FR~I+FR~II/LEG population, the major contribution is provided by
the high radio luminosity, high excitation line, FR~II radio galaxies.

\begin{figure}
\psfig{figure=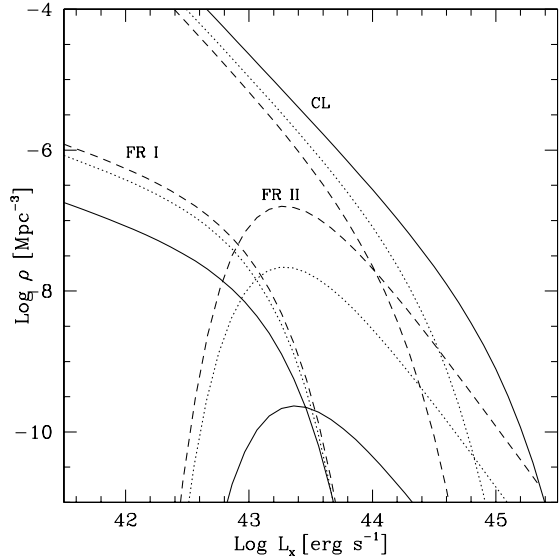,width=0.45\textwidth}
\caption{X--ray luminosity functions of X-ray clusters (CL) and radio
galaxies (low and high radio power populations , labeled for simplicity
FR~I and FR~II, respectively). Continuous, dotted and dashed lines
refer to redshifts z=0, 1, 2, respectively.  The radio 151 MHz
luminosity has been 'converted' into an X-ray (1~keV) one assuming
$\nu_{\rm R} L_{\rm R} = \nu_{\rm x} L_{\rm x}$ and the number density
is per $10^{44}$\ergps.}
\end{figure}

\section{\bf Discussion} 

The above estimates indicate that powerful radio galaxies are expected
to be found in significant numbers at redshift $z\approxgt 1$ as
extended X-ray sources. Indeed, we expect that most old distant radio
galaxies are also extended X--ray sources.

It should be stressed that for any $B< B_{\rm x}$ the above estimates
provide a {\em lower} limit on both the individual luminosity and the
number density of extended X--ray emitting radio sources for the
following reasons. Firstly, extended X-ray emission can also be
produced by non-thermal particles which do not contribute to the 151
MHz emission, further increasing the actual X-ray luminosity with
respect to the estimates given above: lobes, cocoons, relics and jets
can emit not only as inverse Compton emission on the CMB, but also via
other emission processes such as synchrotron self-Compton,
bremsstrahlung and inverse Compton on other photon fields, such as
far-infrared photons in the vicinity of the massive, extremely
luminous galaxies detected in the sub-mm at high redshifts. Secondly,
the radio-emitting electrons cool faster than the X-ray emitting ones
(for $B< B_{\rm x}$). Thus the radio luminosity function might
significantly underestimate, by a factor corresponding to the relative
cooling times $t_{\rm cool} (\gamma_{\rm R})/t_{\rm cool} (\gamma_{\rm
x})$, the number density of sources and the volume pervaded by
non-thermal electrons with energy $\sim \gamma_{\rm x}$ contributing
to the X-ray emission.  In other words the estimates above refer to
'prompt' X-ray emission only over the cooling timescale due to radio
emission. Taking into account this ratio effectively increases the
normalization of the luminosity function of radio sources by factors
$\sim 3-10$ (see Fig. 1b).  In this respect it is interesting to
notice that indeed in the case of 3C294 (Fabian et al 2001) the X-ray
emission extends much further than the radio structure (visible at 5
GHz), indicating that $B<B_{\rm x}$ there.  Consequently, extended
non-thermal X-ray emission is not necessarily associated with a
currently active radio source, instead providing information on the
past radio behaviour of a galaxy.

We conclude that deep X--ray surveys should detect a significant
population of extended X-ray sources associated with both `live and
dead' radio sources. At redshift $z\approxgt 1$ their space density
should be at least comparable to or possibly larger than that of high
X--ray luminosity, high redshift clusters.  Caution in the
interpretation of the origin of extended X--ray emission of course
applies to that associated with high redshift radio sources, treated
as beacons for clusters.  While the presence of a radio source could
still by itself be a cluster tracer (the radio-emitting plasma is
probably confined by some intracluster or intragroup gas), the
inferred luminosities could lead to misleading results on the cluster
luminosity function and thus evolution.  Particularly ambiguous
situations for disentangling X--ray emission from cluster gas or a
present/past AGN activity can arise when only inverse Compton emission
can be detected at the location of a cluster, or when the surface
brightness of inverse Compton emission corresponds to that of the
surrounding intracluster medium\footnote{Powerful radio galaxies and
quasars at $z\sim 0.5-1$ appear to be associated with only poor to
moderately rich clusters with $L_{\rm x}\sim 10^{44}$\ergps (Crawford
\& Fabian 2003; Worrall 2002)}.

Spectral information, including evidence for a thermal component
and/or an iron line feature, such as detected in the cluster RDCS
1252.9-2927 at $z=1.24$ (Rosati et al 2004) and RXJ1053.7+5735 at
$z=1.14$ (Hashimoto et al 2004), will be a key to disentangling the
thermal and non-thermal contributions. The search for extended X--ray
sources in deep $Chandra$ fields (Bauer et al 2002) indicates a
surface density of about 150 deg$^{-2}$ at soft X--ray fluxes $> 3\
10^{-16}$ erg cm$^{-2}$ s$^{-1}$.  A potential diagnostic of
nonthermal emission is emission at energies $\gg1$ keV is also
expected. Although a study of the effective extension of the particle
distribution requires a knowledge of the acceleration processes at
work as well as its history (see Sarazin 1999), if a source is radio
emitting at 151 MHz, for any given $B$ the spectrum produced via the
scattering of CMB photons extends at least up to $\sim 40 (1+z)
B^{-1}_{-6}$ keV; possibly to a few hundreds MeV (for $\gamma \sim
10^6$). Because of photon redshifting, detection of any emission above
a few keV at high $z$ would be a clear signature of a non-thermal
component, presumably due to inverse Compton emission.  Interestingly,
one out of the six extended X-ray sources detected by Bauer et
al. (2002) has a high temperature with respect to the $L_{\rm x}-T$
cluster correlation.

The inverse Compton emission is a direct result, and also a probe, of
the major non-gravitational energy injection phase of the present day
intracluster medium.  In fact, the radio luminosity of radio galaxies
grossly underestimates the intrinsic power of their jets which can be
1000 or more times greater. The radio-emitting plasma is likely to be
confined by some intracluster or intragroup gas, which is displaced
outward.  The energy dumped into the immediate surroundings of such
sources can be considerable and thereby influence the gaseous
properties of clusters and groups (Ensslin et al 1997, Valageas \&
Silk 1999; Wu, Fabian \& Nulsen 2000). Integrating the radio galaxy
luminosity function over time leads to a comoving energy input of
about $10^{57}$\ergpMpc (Inoue \& Sasaki 2001) which can (pre)heat --
although the precise mechanism is not yet clear -- the intracluster
medium by 1--2~keV per particle, so explaining much of the
non-gravitational scaling behavior of groups and clusters (e.g.
Lloyd-Davies et al 2000).  The intracluster or intragroup medium will
be highly disturbed during the energy injection phase and for up to a
core crossing time (about a Gyr) after. This means that the detection
and interpretation of Sunyaev-Zeldovich signals from high redshift
clusters (Carlstron, Holder \& Reese 2003) may be complicated.

Alternatively, the lack of detection of a large population of
non-thermal extended X-ray emitters would provide interesting
information about the radio source/cluster magnetic field evolution.
This would suggest that the radio emitting particles have lower energy
than the radio emitting ones, indicating $B>B_{\rm x}$, and thus point
to positive evolution in the magnetic field associated with the
diffuse radio emission at higher redshifts, although it would be
difficult to precisely quantify and interpret such result.

\noindent 
\section{\bf Acknowledgments} 
We thank the anonymous referee for helpful criticisms.  The Italian
MIUR and INAF (AC) and the Royal Society (ACF) are thanked for
financial support.

\end{document}